\documentstyle[pre,preprint,aps]{revtex} 
\begin{document} 
\title{{\bf Kinetic Temperatures for a Granular Mixture}} 
\author{Steven R. Dahl and Christine M. Hrenya } 
\address{Department of Chemical Engineering, University of Colorado,\\ 
Boulder, CO 80309} 
\author{Vicente Garz\'o} 
\address{Departamento de F\'{\i}sica, Universidad de Extremadura, E-06071 \\ 
Badajoz, Spain} 
\author{James W. Dufty} 
\address{Department of Physics, University of Florida, Gainesville, FL 32611} 
\date{\today} 
\maketitle 
\begin{abstract} 
An isolated mixture of smooth, inelastic hard spheres supports a homogeneous 
cooling state with different kinetic temperatures for each species. This 
phenomenon is explored here by molecular dynamics simulation of a two 
component fluid, with comparison to predictions of the Enskog kinetic 
theory. The ratio of kinetic temperatures is studied for two values of the 
restitution coefficient, $\alpha =0.95$ and $0.80$, as a function of mass 
ratio, size ratio, composition, and density. Good agreement between theory 
and simulation is found for the lower densities and higher restitution 
coefficient; significant disagreement is observed otherwise. The phenomenon 
of different temperatures is also discussed for driven systems, as occurs in 
recent experiments. Differences between the freely cooling state and driven 
steady states are illustrated. 
\end{abstract} 
 
\draft


\section{Introduction} 
\label{sec1} 
 
The dissipative nature of granular media is captured by an idealized fluid 
of smooth, inelastic hard spheres. When isolated and homogeneous such a 
system rapidly approaches a homogeneous cooling state (HCS) for which all 
time dependence of the distribution function occurs through the temperature. 
The latter, defined in the usual way via the average kinetic energy, decays 
in time (``cooling'') due to the inelastic collisions. The existence of the 
HCS and associated cooling rate is well established for a one component 
system by theory\cite{GS95}, Monte Carlo simulation\cite{BRC96}, and 
molecular dynamics simulation\cite{HOB00}. Recently, it has been shown from 
the Enskog kinetic theory that a mixture of inelastic hard spheres also has 
a HCS under the same conditions \cite{GD99}. The condition that all time 
dependence occurs through the temperature requires that the cooling rates 
for the kinetic temperatures for each species must be the same. It follows 
directly that the kinetic temperatures are different for mechanically 
different species, reflecting a violation of the equipartition theorem valid 
for elastic collisions. A prediction for the ratio of temperatures in a 
binary mixture as a function of mass ratio, size ratio, composition, 
density, and restitution coefficients was obtained from an approximate 
solution to the Enskog equations. The accuracy of this approximate result 
has been recently confirmed by Monte Carlo simulation of the Enskog 
equations \cite{MG02}. 
 
The objective here is to demonstrate the phenomenon of a HCS and two 
temperatures in a broader context by molecular dynamics (MD) simulation for 
a binary mixture of inelastic hard spheres. MD simulation avoids any 
assumptions inherent in the kinetic theory or approximations made in solving 
the kinetic equations. It is shown here that MD simulation supports the 
existence of a HCS for mixtures with different kinetic temperatures for each 
species but a common cooling rate. The dependence of the temperature ratio 
on mechanical properties and state conditions is found to be in good 
agreement with predictions of the Enskog kinetic theory except at high 
density and strong dissipation. In the latter case, significant quantitative 
deviations from the Enskog theory are observed but the concept of a HCS and 
two temperatures is preserved. 
 
The HCS can be given a time independent representation by transformation to 
suitable dimensionless variables \cite{L01,D02}. In this form it is similar 
to the steady state obtained for {\em homogeneously driven} granular fluids. 
The latter are obtained by adding stochastic sources to the kinetic equation 
or MD dynamics to do work on the system that compensates for collisional 
cooling. The resulting homogeneous steady state is qualitatively the same as 
the dimensionless HCS, but the quantitative differences are expected to make 
it closer to {\em locally driven} steady states observed in experiments on 
vibrated granular media. Studies of driven states have been extended to 
mixtures both theoretically \cite{BT02} and experimentally \cite{FM01,WP02}. 
Comparisons of the temperature ratio for the HCS mixture and that for two 
types of homogeneously driven mixtures is given below. Their relationship to 
a locally driven system is discussed also. 
 
The plan of the paper is as follows. In Sec.\ \ref{sec2}, we show that the 
Liouville equation for a binary granular mixture supports a scaling solution 
describing the HCS. A transformation to dimensionless variables allows to 
get the (constant) temperature ratio $\gamma=T_1(t)/T_2(t)$ in terms of the 
parameters of the mixture. An approximate evaluation of the temperature 
ratio can be made from the Enskog kinetic theory, as is shown in Sec.\ \ref%
{sec3}. In Sec.\ \ref{sec4}, the Enskog predictions are compared with those 
obtained from MD simulations. Such a comparison shows a quite good agreement 
for the lower densities considered but significant discrepancies between 
theory and simulation appear for high density and strong dissipation. The 
existence of two temperatures in driven granular mixtures 
\cite{BT02,KH00,CH02} and its possible connection with recent experiments is 
analyzed in Sec.\ \ref{sec5}. Finally, the paper ends in Sec.\ \ref{sec6} 
with a brief discussion on the relevance of the results presented here. 
 
\section{Homogeneous Cooling State for a Mixture} 
\label{sec2} 
 
The system considered is a binary mixture of $N_1$ and $N_2$ smooth hard 
spheres of masses $m_{1}$ and $m_{2}$, and diameters $\sigma_{1}$ and $%
\sigma_{2}$. In general, collisions among all pairs are inelastic and are 
characterized by three constant restitution coefficients $\alpha _{ij}$, 
which can be different for the three types of pair collisions. The state of 
the system at time $t$ is specified by the $N=N_1+N_2$ particle phase space 
density $\rho (\Gamma,t)$ which is a solution to the Liouville equation 
\cite {BDS97}. In all of the following, attention is restricted to spatially 
homogeneous states. In this Section it is further assumed that the system is 
isolated. The properties of primary interest are the overall temperature $
T(t)$ associated with the total kinetic energy, and the partial temperatures  
$T_{i}\left(t\right)$ associated with the kinetic energies of each species. 
They are defined as  
\begin{equation}  \label{2.1} 
T(t)=\sum_{i=1}^{2}x_{i}T_{i}\left( t\right), \quad \frac{3}{2}
N_{i}T_{i}\left( t\right) =\left\langle \sum_{\mu =1}^{N_{i}}\frac{1}{2}
m_{i}v_{\mu }^{2};t\right\rangle. 
\end{equation} 
The brackets denote a phase space average over the state of the system at 
time $t$ and $x_{i}=N_{i}/N$ is the composition. The time dependence of $
T(t) $ and $T_i(t) $ follows from the Liouville equation which gives 
\cite{GD99,BDS97,DG01} 
\begin{equation} 
T^{-1}\partial_{t}T=-\zeta ,\quad T_{i}^{-1}\partial _{t}T_{i}=-\zeta_{i}, 
\label{2.3} 
\end{equation} 
where $\zeta_i$ is the cooling rate associated with the partial temperature $%
T_i$ and $\zeta$ is the total cooling rate. They are given by  
\begin{equation} 
\zeta=\frac{1}{T}\sum_{i=1}^{2}x_{i}T_{i}\zeta_{i},  \label{2.4} 
\end{equation} 
\begin{equation} 
\zeta_{i}=-\frac{m_i}{3n_iT_i}\sum_{j=1}^{2}\int d{\bf v}_{1}v_{1}^{2}\int d%
{\bf v}_{2}\int d{\bf r}_{12}\overline{T}_{ij}\left( {\bf r}_{12}, {\bf v}%
_{1}, {\bf v}_{2}\right) f_{ij}^{(2)}\left( {\bf r}_{12},{\bf v}_{1},{\bf v}%
_{2},t\right).  \label{2.4a} 
\end{equation} 
Here, $n_i$ is the number density of species $i$, ${\bf r}_{12}$ is the 
relative position of the two particles, and $f_{ij}^{(2)}\left( {\bf r}
_{12}, {\bf v}_{1},{\bf v}_{2},t\right) $ are the reduced two particle 
distribution functions for a particle of type $i$ and one of type $j$, 
obtained from $\rho(\Gamma ,t)$ by integrating over degrees of freedom for 
all other particles. The binary collision operators are defined by \cite
{BDS97,DG01}  
\begin{equation} 
\overline{T}_{ij}\left( {\bf v}_{1}, {\bf v}_{2}, {\bf r}_{12}\right) 
=\sigma _{ij}^{2}\int d\widehat{\bbox {\sigma}}\,\Theta (\widehat{\bbox 
{\sigma}}\cdot {\bf g}_{12})(\widehat{\bbox {\sigma }}\cdot {\bf g}_{12}) 
\left[ \alpha_{ij}^{-2}\delta ({\bf r}_{12}-\bbox{\sigma})b_{ij}^{-1}-\delta 
({\bf r}_{12}+\bbox{\sigma})\right]  \label{2.6} 
\end{equation} 
where $\sigma_{ij}=\left( \sigma _{i}+\sigma _{j}\right) /2$, $\widehat{
\bbox {\sigma}}$ is a unit vector directed along the line of centers from 
the sphere of species $i$ to that of species $j$ at contact, $\Theta $ is 
the Heaviside step function, and ${\bf g}_{12}={\bf v}_{1}-{\bf v}_{2}$. 
Also, $b_{ij}^{-1}$ is a substituting operator, $b_{ij}^{-1}F({\bf g} 
_{12})=F\left(b_{ij}^{-1}{\bf g}_{12}\right)$, which changes any function of  
${\bf v}_{1}$ and ${\bf v}_{2}$ to the same function of the restituting 
velocities ${\bf v}_{1}^{\prime }$ and ${\bf v}_{2}^{\prime}$:  
\begin{equation} 
{\bf v}_{1}^{\prime }={\bf v}_{1}-\mu _{ji}\left( 1+\alpha _{ij}^{-1}\right) 
(\widehat{\bbox {\sigma}}\cdot {\bf g}_{12})\widehat{\bbox {\sigma}}, \quad  
{\bf v}_{2}^{\prime }={\bf v}_{2}+\mu _{ij}\left( 1+\alpha _{ij}^{-1}\right) 
(\widehat{\bbox {\sigma}}\cdot {\bf g}_{12})\widehat{\bbox {\sigma}} 
\label{2.7} 
\end{equation} 
where $\mu_{ij}=m_{i}/\left( m_{i}+m_{j}\right)$. Upon writing Eqs.\ (\ref%
{2.4a}) and (\ref{2.6}) we have taken into account that for an homogeneous 
system the spatial dependence of $f_{ij}$ occurs only through ${\bf r}_{12}$. 
 
In general, all three temperatures and associated cooling rates will be 
different and depend on the initial preparation. The time evolution of the 
ratio of the two partial temperatures $\gamma(t)=T_{1}(t)/T_{2}(t)$ follows 
from the second equality of Eq.\ (\ref{2.3}):  
\begin{equation} 
\partial_{t}\ln \gamma =\zeta_{2}-\zeta_{1}.  \label{2.8} 
\end{equation} 
For a system with elastic collisions, $\rho(\Gamma ,t)$ rapidly approaches 
to the Gibbs distribution with a single constant temperature. This requires $%
\zeta_{1}=\zeta_{2}$, and $T_{i}\propto T$ in the Gibbs state. The form of 
the velocity distribution functions and the constancy of the temperature 
then gives $T_1=T_2=T$ and $\zeta_1=\zeta_2=0$. This equality of the 
temperatures is the equipartition theorem for classical statistical 
mechanics. The vanishing of the cooling rates is a consequence of the system 
approaching towards a steady state.

If the collisions are inelastic, the system still approaches rapidly a 
special state known as the homogeneous cooling state (HCS). As with the 
Gibbs state, the velocities scale with the temperature for a dimensionless 
universal distribution of the form  
\begin{equation} 
\rho_{hcs}(\Gamma ,t)=\left[\ell v(t)\right] ^{-3N}\rho _{hcs}^{\ast }\left( 
\{{\bf r}_{ij}^{\ast},{\bf v}_{i}^{\ast}\}\right) .  \label{2.9} 
\end{equation} 
Here ${\bf r}_{ij}^{\ast}={\bf r}_{ij}/\ell $ denotes the dimensionless 
relative coordinate for particles $i$ and $j$, and $\ell $ is some 
appropriate characteristic length scale such as the mean free path. The 
dimensionless velocities ${\bf v}_{i}^{\ast}={\bf v}_{i}/v_{0}(t)$ are 
scaled relative to the thermal velocity defined by $v_{0}(t)= \sqrt{
2T(t)\left( m_{1}+m_{2}\right) /m_{1}m_{2}}$. This scaling has the same 
consequences as described above for elastic collisions: $T_{i}(t)\propto 
T(t) $, $\gamma(t) \rightarrow \text{constant}$, and so 
$\zeta_{1}=\zeta_{2}$. However, Eq.\ (\ref{2.9}) is not a steady state since 
the cooling rates $\zeta_i$ do not vanish. Also, the form of 
$\rho_{hcs}^{\ast}$ is not the same as for the Gibbs state so there is no 
{\em a priori} reason to expect that the temperatures should be 
equal\cite{GD99}. In fact, as indicated below, they are equal only in the 
limit of mechanically equivalent particles or elastic collisions. 
 
The simplest test of the evolution towards a HCS with the assumed velocity 
scaling is the approach of $\gamma \left( t\right) $ to a constant value. 
This is illustrated in Fig.\ \ref{fig1} from MD simulation using $1000$ 
particles of the same size and composition, but with the mass ratio $
m_{1}/m_{2}=8$. The total solid volume fraction is $\phi=0.1$ and all 
coefficients of restitution are equal. Here, $\phi=\phi_1+\phi_2$, where 
\begin{equation}
\label{2.9bis}
\phi_i=\frac{1}{6}\pi n_i \sigma_i^3
\end{equation}
is the species volume fraction of the component $i$. We consider two values 
of the restitution coefficient: $ \alpha=0.8$ and $\alpha=1$. We observe 
that in the elastic case ($\alpha=1$) the mixture approaches the Gibbs state 
with the temperature ratio $%
\gamma(t)\to 1$, as expected from equipartition. In the inelastic case ($%
\alpha=0.8$), $\gamma(t)$ approaches a constant value ($\gamma \simeq 2$ 
with fluctuations less than $5\%$) after about $10$ collisions per particle. 
It is seen that the the HCS for inelastic collisions is approached on the 
same time scale as the Gibbs state for elastic collisions. Further details 
of the MD simulation are discussed below. 
 
\section{Enskog Kinetic Theory} 
\label{sec3} 
 
The kinetic temperatures defined by Eq.\ (\ref{2.1}) can be given in an 
equivalent form in terms of the one particle reduced distribution function $%
f_{i}^{(1)}({\bf v},t)$ as  
\begin{equation} 
T_{i}\left( t\right) =\frac{1}{3}\frac{m_{i}}{n_i}\,\int \, d{\bf v}\, 
v^{2}\,f_{i}^{(1)}({\bf v},t).  \label{3.1} 
\end{equation} 
The one particle reduced distribution function $f_{i}^{(1)}({\bf v},t)$ 
obeys the exact first BBGKY hierarchy equations  
\begin{equation} 
\partial_{t}f_{i}^{(1)}({\bf v}_{1},t)=\sum_{j=1}^{2}\int d{\bf v}_{2}\int d%
{\bf r}_{12}\overline{T} _{ij}\left({\bf r}_{12}, {\bf v}_{1}, {\bf v}%
_{2}\right) f_{ij}^{(2)} \left({\bf r}_{12}, {\bf v}_{1},{\bf v}_{2}, 
t\right) .  \label{3.2} 
\end{equation} 
To be more specific about the dependence of the temperatures on the 
parameters of the mixture it is sufficient to specify the reduced 
distribution functions $f_{ij}^{(2)}\left({\bf r}_{12}, {\bf v}_{1}, {\bf v} 
_{2}, t\right) $ in Eq.\ (\ref{3.2}). This also determines the cooling rates 
from Eq.\ (\ref{2.4a}). These distribution functions occur only in the 
combination $\overline{T}_{ij}f_{ij}^{(2)}$, so knowledge of $f_{ij}^{(2)}$ 
is required only for pairs of particles at contact and only on the 
pre--collision hemisphere. A practical approximation for these conditions is 
obtained by neglecting velocity correlations and expressing the two particle 
distribution functions in terms of the single particle distribution 
functions 
\begin{equation} 
f_{ij}^{(2)}\left({\bf r}_{12}, {\bf v}_{1},{\bf v}_{2}, t\right) 
\rightarrow f_{i}^{(1)}\left( {\bf v}_{1},t\right) f_{j}^{(1)}\left( {\bf v}
_{2},t\right) \chi_{ij}\left({\bf r}_{12},t\right) .  \label{3.2a} 
\end{equation} 
The single particle distributions are independent of position since only \ 
homogeneous states are considered here. The spatial correlation function $%
\chi _{ij}\left( {\bf r}_{12},t\right) $ is evaluated at contact and its 
choice is given below. The one particle distribution functions can be 
determined by using this same approximation in the exact first BBGKY 
hierarchy equations, which becomes  
\begin{equation} 
\partial _{t}f_{i}^{(1)}({\bf v}_{1},t)=\sum_{j=1}^2J_{ij}\left[{\bf v} 
_{1}|f_{i}^{(1)}(t),f_{j}^{(1)}(t)\right] \;,  \label{3.3} 
\end{equation} 
where $J_{ij}[f_{i}^{(1)},f_{j}^{(1)}]$ is the Enskog collision 
operator\cite {BDS97}. These are now closed equations for $f_{i}^{(1)}$ and 
constitute the Enskog kinetic theory for the granular binary mixture. 
 
For the HCS the scaling form (\ref{2.9}) implies a similar scaling form for $%
f_{i}^{(1)}({\bf v}_{1},t)$  
\begin{equation} 
f_{i}^{(1)}({\bf v},t)=n_{i}v_{0}^{-3}(t) f_{i}^*\left(v^{\ast}\right) . 
\label{3.4} 
\end{equation} 
Furthermore, $\chi_{ij}\left({\bf r}_{12}=\sigma _{ij},t\right) \rightarrow 
\chi_{ij}\equiv \text{constant}$ since all time dependence occurs through 
the velocity scaling. For practical purposes, and to agree with the 
equilibrium limit for elastic collisions, $\chi_{ij}$ is taken to be the 
equilibrium pair correlation function. A good approximation is given by the 
Carnahan--Starling form \cite{GH72}  
\begin{equation} 
\chi_{ij}=\frac{1}{1-\phi}+\frac{3}{2}\frac{\xi}{(1-\phi)^{2}} \frac{ 
\sigma_{i}\sigma_{j}}{\sigma_{ij}}+\frac{1}{2} \frac{\xi^{2}}{(1-\phi)^{3}} 
\left(\frac{\sigma_{i}\sigma_{j}}{\sigma_{ij}}\right)^{2}\;,  \label{3.4a} 
\end{equation} 
where $\xi=\pi\left(n_{1}\sigma_{1}^{2}+n_{2}\sigma _{2}^{2}\right)/6$. 
Comparison with computer simulations for binary molecular hard sphere 
mixtures have shown that the Carnahan--Starling equation (\ref{3.4a}) is 
accurate in most of the fluid region, although it fails for high densities 
and for large diameter ratios\cite{CS}. Given the values considered in our 
simulations, we expect that the approximation (\ref{3.4a}) turns out to be 
quite accurate to evaluate the pair correlation function $\chi_{ij}$. In 
terms of the reduced distributions $f_i^*$, the Enskog kinetic equations 
become  
\begin{equation} 
\frac{\zeta_{i}^{\ast }}{2}\frac{\partial}{\partial {\bf v}^{\ast }}\cdot 
\left( {\bf v}^{\ast }f_{i}^*\right) =\sum_{j=1}^2J_{ij}^{\ast}\left[ {\bf v}%
^{\ast }|f_{i}^*,f_{j}^*\right],  \label{3.5} 
\end{equation} 
where $\zeta _{i}^{\ast }=\zeta _{i}/nv_{0}\sigma _{12}^{2}$ and $%
J_{ij}^{\ast }=\left(v_{0}^{2}/nn_{i}\sigma _{12}^{2}\right) J_{ij}$ are 
given, respectively, by  
\begin{equation}  \label{3.5a} 
\zeta_i^*[f_i^*,f_j^*]=-\frac{2}{3}\lambda_i\sum_{j=1}^2\,\int\, d{\bf v}%
_1^* v_1^{*2}J_{ij}^*[f_i^*,f_j^*], 
\end{equation} 
\begin{eqnarray}  
\label{3.5b} 
J_{ij}^*[v_1^*|f_{i}^*,f_{j}^*]&=&x_j\chi_{ij}\left(\frac{\sigma_{ij}} {%
\sigma_{12}}\right)^2\int d{\bf v}_2^* \int d\widehat{\bbox {\sigma}}%
\,\Theta (\widehat{\bbox {\sigma}}\cdot {\bf g}_{12}^*)(\widehat{\bbox 
{\sigma }}\cdot {\bf g}_{12}^*)  \nonumber \\ 
& & \times \left[ \alpha_{ij}^{-2}f_i^*(v_1^{\prime*})f_j^*(v_2^{
\prime*})-f_i^*(v_1^{*}) f_j^*(v_2^{*})\right], 
\end{eqnarray} 
where ${\bf g}_{12}^*={\bf g}_{12}/v_0$, $\lambda_{i}=\left(v_{0}/v_{0i}
\right) ^{2}=T/(T_i\mu_{ji})$, with $j\neq i$, is the square of the 
ratio of the thermal velocity to that for species $i$, and  
$v_{0i}=\sqrt{2T_{i}/m_{i}}$.

In the dimensionless form (\ref{3.5}) the Enskog equations are time 
independent. The pair of coupled equations (\ref{3.5}) must be solved 
self-consistently with the expressions for the cooling rates in Eq.\ (\ref%
{2.4a}) to determine $f_{i}^*$ and $\zeta_{1}^{\ast}=\zeta_{2}^{\ast}=%
\zeta^{\ast}$. The temperature $T(t)$ is then obtained from the known 
cooling rate by solving the first of Eqs.\ (\ref{2.3}), and the distribution 
functions $f_{i}^{(1)}({\bf v},t)$ are fully determined. The kinetic 
temperature for each species is obtained from Eq.\ (\ref{3.5}) as  
\begin{equation} 
T_{i}(t) =T(t)\frac{2}{3}\mu_{ji}^{-1}\int d{\bf v}^{\ast }v^{\ast 
2}f_i^*\left( v^{\ast}\right) .  \label{3.6} 
\end{equation} 
As anticipated, $T_{i}(t) \propto T(t)$ in the HCS and $\gamma=T_1/T_2$ 
becomes  
\begin{equation} 
\gamma=\frac{\mu_{12}}{\mu_{21}}\frac{\int d{\bf v}^{\ast}v^{\ast 2}f 
_{1}^*\left( v^{\ast }\right)}{\int d{\bf v}^{\ast }v^{\ast 2}f 
_{2}^*\left(v^{\ast }\right)}.  \label{3.7} 
\end{equation} 
This is essentially the approach used in numerical Monte Carlo solution \cite%
{MG02} to the Enskog equation. 
 
In practice, only approximate solutions for the HCS are possible (an 
exception is a recent exact result for a one dimensional Maxwell model \cite%
{MP02}) and a different approach is followed. First, the solution is 
represented as a series in velocity polynomials, with the leading terms 
given by  
\begin{equation} 
f_{i}^*(v^*)\rightarrow \left( \frac{\lambda_{i}}{\pi}\right) 
^{3/2}e^{-\lambda _{i}v^{*2}}\left[ 1+\frac{c_{i}}{4}\left( \lambda 
_{i}^{2}v^{* 4}-5\lambda_{i}v^{* 2}+\frac{15}{4}\right) \right]  \label{3.8} 
\end{equation} 
Thus, the weight function (Gaussian) for each species is chosen to be scaled 
relative to the thermal velocity for that species, introducing explicitly 
the unknown species temperatures. The coefficients $c_i$ measure the 
deviation of $f_i^*$ from the chosen reference Gaussians. The cooling rates 
are now calculated as explicit functions of $\lambda_{i}$ and $c_{i}$ from 
Eq.\ (\ref{3.5a}). With these known, the Enskog equations can be solved to 
determine $c_{i}$ as functions of $\lambda _{i}$ by substitution of (\ref 
{3.8}) into the Enskog equations, taking the $v^{4}$ moment of those 
equations, and retaining terms up through linear in $c_{i}$. Finally, the $ 
\lambda _{i}$ are determined from the consistency condition for the HCS, $ 
\zeta_{1}^*=\zeta_{2}^*$. The detailed results for $c_{i}$ and $\lambda_{i}$ 
as functions of the fluid parameters are given in Refs.\ \cite{GD99} and  
\cite{MG02} and will not be repeated here.

\section{Comparison of Theory and Simulation} 
\label{sec4} 
 
The approximation (\ref{3.8}) provides detailed predictions for the species 
temperatures as functions of the mass ratio, size ratio, composition, 
density, and restitution coefficients. The quality of this approximate 
solution to the Enskog equations has been recently confirmed by direct Monte 
Carlo simulation of those equations over a wide range of the parameter space 
\cite{MG02}. Specifically, the parameter space over which the solution 
(\ref{3.8}) has been verified is the mass ratio $m_1/m_2$, the 
concentration ratio $n_1/n_2$, the ratio of diameters $\sigma_1/\sigma_2$, 
the reduced density $n\sigma_{12}^3$, and the (common) restitution 
coefficient $\alpha\equiv \alpha_{11}=\alpha_{22}=\alpha_{12}$. However, 
uncertainties remain regarding the accuracy of the Enskog equations 
themselves. An appropriate means to study the concept of the HCS and 
associated different partial temperatures, as well as to study the domain of 
validity of the Enskog kinetic theory is via MD simulations. Since the 
parameter space here is quite large the tests of the theory and concepts are 
quite stringent. 
 
Two different values of the solid volume fractions $\phi$ have been 
considered here, $\phi=0.1$ and $\phi=0.2$, both representing a moderately 
dense fluid. All coefficients of restitution were set equal and two values 
considered, $\alpha =0.8$ and $\alpha =0.95$, both representing moderately 
strong dissipation. The temperature ratio $\gamma$ in the HCS has been 
studied for three cases in each state. In the first case (case I) $\gamma$ 
is determined as a function of the mass ratio $m_{1}/m_{2}$ for $\sigma 
_{1}/\sigma _{2}=\phi_{1}/\phi_{2}=1$. The second case (case II) determines $%
\gamma$ as a function of size ratio $\sigma_{1}/\sigma_{2}$ for $%
m_{1}/m_{2}=\phi_{1}/\phi_{2}=1$, while the third case (case III) determines  
$\gamma$ as a function of composition $\phi_{1}/\phi_{2}$ for $m_{1}/m_{2}=8$ 
and $\sigma_{1}/\sigma_{2}=2$.

The granular system under consideration does not contain external force 
fields, and thus the particles travel in straight-line trajectories between 
collisions. Correspondingly, an event-driven algorithm is employed in the MD 
simulations. The simulated particles are modeled as inelastic, frictionless 
hard spheres (i.e. collisions are both binary and instantaneous) moving in a 
three-dimensional space with standard periodic boundaries. The initial 
particle velocities are uniformly distributed about a zero mean, regardless 
of particle size. These velocities are then adjusted to ensure that the net 
system momentum is zero.

As indicated in Fig.\ \ref{fig1}, the system reaches a steady value for 
the temperature ratio $\gamma$ within 10 collisions per particle for a wide 
class of initial conditions. However, it is known that 
the HCS is unstable to long wavelength perturbations so that spontaneous 
deviations from the HCS occur at long times. To assure that $\gamma $ is 
measured in the HCS, the temperature $T(t)$ is monitored as a function of 
time to determine if the predicted cooling from the scaling form (\ref{2.9}) 
(Haff's law) is maintained\cite{footnote}. In order to keep the 
computational time reasonable for each of the simulations (about 1 hour), 
the total number of particles was kept constant at $N=1000$ for all 
simulations. Data from the first 10 collisions per particle (or 10000 
total collisions) was used to determine the slope of $T(t)$. 
Specifically, the $\ln (T(t)/T(0))$ was sampled $1000$ times during 
this initial portion of the simulation. Somewhat surprisingly, a smooth 
linear decrease in $\ln (T(t)/T(0))$ was observed throughout the 1000 
samples, and a linear regression analysis was employed to evaluate the slope 
of the Haff's law plot. Following evaluation 
of the slope of $\ln (T(t)/T(0))$, collection of the $\gamma 
=T_{1}(t)/T_{2}(t)$ data commenced. This data collection period involved as 
many as $200$ additional collisions per particle (or $200000$ total 
collisions) including $50000$ equally spaced measurements of $\gamma $. The 
phrase ``as many as'' refers to the fact that the data collection would 
cease (with fewer than $50000$ measurements of the energy ratio) if the 
measured value of $\ln (T(t)/T(0))$ deviated from the expected value of $\ln 
(T(t)/T(0))$ by more than $5\%$. Violation of the Haff's law 
restriction occurred frequently when the mass ratio $m_1/m_2$ was greater 
than 4. Additionally, simulations of equal mass particles ($m_1/m_2=1$) 
violated the Haff's law restriction when $\alpha=0.8$ and $\phi=0.2$.

Figure \ref{fig2} shows the results for case I, $\gamma$ as a function of 
mass ratio. The symbols represent the simulation data where the circles are 
for $\alpha=0.95$ and the triangles are for $\alpha=0.8$. In addition, open 
(solid) symbols correspond to $\phi=0.1$ ($\phi=0.2$). The simulation values 
reported represent the average from three identical simulations, with a 
standard deviation typically less than $3\%$. The Enskog prediction of the 
previous section is given by the solid lines (the theory does not predict 
any dependence on $\phi$ in this case). The agreement between the theory and 
simulation is seen to be quite good at $\alpha=0.95$, over the whole range 
of mass ratios. The agreement is also quite good at $\alpha=0.8$ and $%
\phi=0.1$. However, systemmatic deviations from the Enskog theory for large 
mass ratios are obtained in the simulations at $\phi=0.2$. 
 
Figure \ref{fig3} shows the results for case II, $\gamma$ as a function of 
size ratio. The notation is the same as in Fig. \ref{fig2} where now the 
solid line refers to $\phi=0.1$ while the dashed line is for $\phi=0.2$. The 
agreement for both $\alpha=0.95$ and $\alpha=0.8$ is quite good at $\phi=0.1$%
, except for the largest size ratio at $\alpha=0.8$. The density dependence 
of the theory is weaker than that from the simulation, and large differences 
are observed at $\phi=0.2$. 
 
Figure \ref{fig4} shows the results for case III, $\gamma$ as a function of 
composition. We observe that both the theory and simulation predicts a very 
weak influence of composition on the temperature ratio. In addition, the 
trends are similar to those of Figs.\ \ref{fig2} and \ref{fig3}. Good 
agreement is obtained for $\alpha=0.95$ at both $\phi=0.1$ and $\phi=0.2$. 
At stronger dissipation there is a strong density dependence in the 
simulation that is not reproduced by the theory.

\section{Driven Systems and Experiments} 
\label{sec5} 
 
The existence and details of different temperatures for each species in a 
HCS is now well established by kinetic theory and simulation. Related 
experiments \cite{FM01,WP02} and simulations \cite{KH00,CH02} on driven 
steady states also show different temperatures, but the detailed dependence 
on the control parameters appears to be different. The driven steady states 
are achieved from external forces that do work at the same rate as 
collisional cooling. In the experiments this is accomplished by vibrating 
the system so that it is locally driven at the walls. Far from these walls a 
steady state is studied whose properties are presumed to be insensitive to 
the details of the driving forces. The velocities of the particles can be 
measured using high speed photography 
\cite{FM01} or positron emission particle tracking \cite{WP02}. The 
objective of this section is to explore similarities and differences between 
the temperature ratios for a binary mixture in the HCS and in a driven 
steady state. 
 
As a first analysis, a {\em homogeneously} driven steady state is 
considered. This does not correspond directly to any experimental driving 
source but has been considered extensively as a representation of driven 
systems for the one component fluid\cite{thermostat}. In this case a uniform 
external non-conservative force, frequently referred to as a ``thermostat'', 
is applied to compensate for collisional cooling. Two types of thermostats 
are considered here. One is a deterministic Gaussian thermostat widely used 
in nonequilibrium molecular dynamics simulation for molecular fluids\cite%
{EM90}. The force is similar to a Stokes law drag force, linear in the 
velocity, but with the opposite sign so that it heats rather than cools the 
system. The ``friction'' constant can be chosen to exactly compensate for 
collisional cooling. At the level of kinetic theory, the introduction of 
such an external force leads to a steady state equation which is {\em  
identical} to Eq.\ (\ref{3.5}). It is easily confirmed that the same is true 
at the level of the Liouville equation in appropriate dimensionless 
variables. Thus, there is an exact correspondence between the HCS and this 
type of driven steady state and, in particular, the dependence of $\gamma $ 
on the control parameters is the same. 
 
A second method of driving the system homogeneously is by means of a 
stochastic Langevin force representing Gaussian white noise \cite{DM96}. 
This force for each species is written as ${\cal F}_{i}=m_{i}\bbox{\xi}_{i}$%
, where the covariance of the stochastic acceleration is  
\begin{equation} 
\langle\xi_{i\alpha}(t) \xi_{j\beta}(t^{\prime})\rangle 
=2D\delta_{ij}\delta_{\alpha \beta}\delta \left(t-t^{\prime}\right) 
\label{5.1} 
\end{equation} 
Note that the covariance for the random accelerations is taken to be the 
same for each species\cite{BT02,HBB00}. This force induces a diffusion in 
velocity space, with diffusion coefficient $D$. At the level of kinetic 
theory this leads to an additional source represented by a Fokker--Planck 
collision operator in addition to the Enskog collision operator. The steady 
state Enskog equations then take the form  
\begin{equation} 
0=\sum_jJ_{ij}\left[{\bf v}|f_{i}^{(1)},f_{j}^{(1)}\right]+D \left(\frac{%
\partial}{\partial v}\right)^2f_{i}^{(1)}.  \label{5.2} 
\end{equation} 
Multiplying by $m_iv_{i}^{2}/2$ and integrating gives the relationship of 
$D$ to the cooling rates $\zeta_i$, i.e., $D=\zeta _{i}T_{i}/2m_{i}$. This 
in turn implies the steady state condition  
\begin{equation} 
\zeta _{1}\frac{T_{1}}{m_{1}}=\zeta _{2}\frac{T_{2}}{m_{2}}  \label{5.3} 
\end{equation} 
The cooling rates are no longer equal, as for the HCS, and the dependence of 
the temperatures on the control parameters will be different as well. 
 
The procedure for determining the temperatures for the stochastically driven 
steady state is the same as that described in Sec.\ \ref{sec3}. The steady 
state distribution is represented as an expansion of the form (\ref{3.8}) 
and the coefficients are now determined from moments of the set (\ref{5.2}). 
The cooling rates are then determined from this solution using Eq.\ (\ref%
{3.5a}), and the condition (\ref{5.3}) gives an equation for the temperature 
ratio $\gamma$. Figures \ref{fig5}, \ref{fig6}, and \ref{fig7} illustrate 
the differences between the HCS and the stochastic steady state for $\alpha 
=0.6$ and $0.8$. The solid lines are the results for the HCS while the 
dashed lines are the results for the driven steady state. The dependence of $%
\gamma$ on mass ratio is shown in Fig.\ \ref{fig5} for $\phi=0$ and $%
\phi_{1}/\phi_{2}=$ $\sigma_{1}/\sigma_{2}=1$. This dependence is seen to be 
considerably stronger in the driven state. The dependence on composition is 
shown in Fig.\ \ref{fig6} for $\phi=0$, $m_{1}/m_{2}=2$ and $%
\sigma_{1}/\sigma_{2}=1$. Finally the dependence on overall packing fraction  
$\phi$ is shown in Fig.\ \ref{fig7} for $\phi_{1}/\phi_{2}=1$ and $%
m_{1}/m_{2}=\sigma_{1}/\sigma_{2}=2$. In this last case the effect of 
increased density is greater for the HCS than for the driven steady state. 
 
The HCS and homogeneously driven steady states are seen to be qualitatively 
similar, with only quantitative differences. It remains to understand their 
relationship to locally driven wall forces. An example is described for the 
Boltzmann equation in the Appendix. There, the boundary condition is a 
sawtooth vibration of one wall such that every particle encountering the 
wall has a reflected speed increased by twice the velocity of the wall in 
the component normal to the wall. The steady state condition is considerably 
more complex than for the HCS or the homogeneously driven steady states. In 
the limit that the wall velocity is large compared to the thermal velocities 
of each species the condition (\ref{5.3}) is recovered. This suggests that 
the results obtained from this condition are plausible first approximations 
for qualitative comparisons with experimental results \cite{BT02}. However, 
the detailed nature of the driven state requires further characterization 
before quantitative conclusions can be drawn. This is suggested by the study 
of a driven state in the absence of gravity \cite{BRM00} where the system is 
found to be well-described by hydrodynamics away from the wall, but the 
steady state is strongly inhomogeneous.

\section{Discussion} 
\label{sec6} 
 
The primary results of this study are two-fold. First, the MD simulations 
confirm the rapid approach to a HCS with two kinetic temperatures determined 
by a common cooling rate. This occurs over a wide range of densities, 
composition, mass and size ratios, for both moderate and strong dissipation. 
The second result is confirmation of the Enskog kinetic theory to provide a 
quantitative description of this phenomenon for the lower densities and 
weaker dissipation cases. This includes densities well outside the Boltzmann 
limit and applies throughout the parameter space of mechanical properties. 
The analysis here is a test of the Enskog prediction for the cooling rates, 
which are essentially transport properties (collision rates). The good 
agreement obtained is further testimony to the utility of this remarkable 
equation for fluids with elastic and inelastic collisions, including 
mixtures. 
 
The failure of the Enskog theory at high densities is expected from 
experience with normal fluids. This is due to multi-particle collisions, 
including recollision events (ring collisions). The latter are expected to 
be stronger for fluids with inelastic collisions where colliding pairs tend 
to become more focused. It appears that the range of densities for which the 
Enskog description applies decreases with increasing dissipation. This is 
the case observed here and also elsewhere for the self-diffusion coefficient %
\cite{D02}. The specific mechanism responsible for the large discrepancies 
at high densities and its quantitative prediction remains an open problem. 
 
The magnitude of the difference between the two kinetic temperatures 
generally increases as the mechanical differences increase, although the 
dependence on volume fraction is weak. Also, there is a significant 
dependence on the inelasticity and total volume fraction. The experiments in %
\cite{FM01} show a similar strong dependence on mass ratio, but no 
significant dependence on inelasticity, total density, or composition. The 
detailed correspondence between the simple model homogeneous states 
considered here and the locally driven states of experiments needs 
refinement, although generally the same trends are observed \cite{BT02}. 
 
The hydrodynamics for binary mixtures of inelastic hard spheres has been 
derived recently, including the effects of two kinetic temperatures \cite%
{GD02}. Although only the overall temperature associated with both \ species 
serves as a hydrodynamic field, the transport coefficients depend on the 
temperature ratio $\gamma $. Since the latter is a function of the 
composition and density, there are additional contributions to the transport 
coefficients. Differences as large as $50\%$ are found for some coefficients.

\acknowledgements 
S. R. D. is grateful for the support provided by the National Science 
Foundation Graduate Fellowship Program. V.G. acknowledges partial support 
from the Ministerio de Ciencia y Tecnolog\'{\i}a (Spain) through Grant No. 
BFM2001-0718.

\appendix  
\section{Local boundary conditions} 
 
The Enskog equations with boundary conditions can be written as \cite{DvB77}
\begin{equation} 
\left(\partial_{t}+{\bf v}_1\cdot\nabla -T_{Wi}\right) f_{i}^{(1)}({\bf r},
{\bf v}_{1};t)=\sum_{j}J_{ij}\left[{\bf v}_{1}|f_{i}^{(1)}(t),f_{j}^{(1)}(t) 
\right] \;,  \label{a.1} 
\end{equation}
Here, $T_{Wi}$ describes interactions of particles of type $i$ with the 
boundaries  
\begin{eqnarray} 
T_{Wi}f_{i}^{(1)}({\bf r},{\bf v}_{1};t) &=&
\int_{\Xi}d{\bf s}\delta \left( 
{\bf r}-{\bf s}\right) \left\{\int d{\bf v}^{\prime}
P({\bf v}_{1},{\bf v}^{\prime})\left|\widehat{{\bf n}}\cdot {\bf 
v}^{\prime }\right| f_{i}^{(1)}({\bf r},{\bf v}^{\prime };t)\right.  
\nonumber \\ &&\left. -\Theta (-\widehat{{\bf n}}\cdot {\bf v}_{1})\left| 
\widehat{{\bf n}}\cdot {\bf v}_{1}\right| f_{i}^{(1)}({\bf r},{\bf 
v}_{1};t)\right\} \label{a.2} 
\end{eqnarray}
This is similar to the usual Boltzmann collision operator with the first 
term representing a gain in the population of particles with velocity 
${\bf v}_{1}$ due to collisions with the wall and the second term 
representing a corresponding loss. The probability density for a velocity 
${\bf v}_{1}$ after the collision with the wall, given an incident velocity 
${\bf v}^{\prime }$ has the form  
\begin{equation} 
P({\bf v}_{1},{\bf v}^{\prime})=\Theta (\widehat{{\bf n}}\cdot {\bf v} 
_{1})K({\bf v}_{1},{\bf v}^{\prime })\Theta (-\widehat{{\bf n}}\cdot {\bf v} 
^{\prime }).  \label{a.3} 
\end{equation}
The two theta functions characterize incident particles as directed toward 
the wall and reflected particle directed away from the wall, where the 
normal $\widehat{{\bf n}}$ is directed toward the interior of the system. 
The kernel $K({\bf v}^{\prime },{\bf v}_{1})$ characterizes the change in 
the half space velocity distributions at the surface (i.e., outgoing 
distribution is a linear functional of the incoming distribution). Particle 
number conservation requires  
\begin{equation} 
\int d{\bf v}_{1}P({\bf v}_{1},{\bf v}^{\prime })=\Theta (-\widehat{{\bf n}}%
\cdot {\bf v}^{\prime })  \label{a.4} 
\end{equation} 
As an illustration, the form of $K({\bf v},{\bf v}_{1})$ for elastic 
specular collisions with a wall at rest is  
\begin{equation} 
K_{s}({\bf v}_{1},{\bf v}^{\prime })=\delta \left( {\bf v}_{1}-{\bf v}
^{\prime }+2\left( \widehat{{\bf n}}\cdot {\bf v}^{\prime }\right) \widehat{
{\bf n}}\right)  \label{a.5} 
\end{equation} 
 
The surface domain $\Xi $ is time dependent, representing vertical 
oscillations. This can be handled using a sawtooth form for the driving, 
such that every particle encounters the wall moving into the system at 
velocity ${\bf v}_{w}=v_{w}\widehat{{\bf n}}$. Assume the amplitude of the 
vibration is small so that the displacement of the wall can be neglected. 
Then, it is reasonable to choose specular collisions in the local Galillean 
frame for which the wall is at rest:  
\begin{equation} 
P({\bf v}_{1},{\bf v}^{\prime })=\Theta (\widehat{{\bf n}}\cdot \left( 
{\bf v}_{1}{\bf -v}_{w}\right) )K({\bf v}_{1},{\bf v}^{\prime })\Theta 
(-\widehat{ 
{\bf n}}\cdot \left( {\bf v}^{\prime }-{\bf v}_{w}\right)),  \label{a.6} 
\end{equation}
\begin{equation} 
K_{s}({\bf v}_{1},{\bf v}^{\prime })=\delta \left( {\bf v}_{1}-{\bf v} 
^{\prime }+2\left( \widehat{{\bf n}}\cdot \left( {\bf v}^{\prime}-{\bf v} 
_{w}\right) \right) \widehat{{\bf n}}\right).  \label{a.6a} 
\end{equation} 
The wall collision operator becomes  
\begin{eqnarray} 
T_{Wi}f_{i}^{(1)}({\bf r},{\bf v}_{1};t) &=&
\int_{\Xi}d{\bf s}\delta \left( 
{\bf r}-{\bf s}\right) \left\{ \Theta (\widehat{{\bf n}}\cdot \left( {\bf 
v}_{1}-{\bf v}_{w}\right))\int d{\bf v}^{\prime }\delta \left( {\bf 
v}_{1}-{\bf v}^{\prime }+2\widehat{{\bf n}}\cdot\left({\bf v}^{\prime 
}-{\bf v}_{w}\right) \widehat{{\bf n}}\right) \right.  \nonumber \\ 
& &
\times \Theta (-\widehat{{\bf n}}\cdot \left({\bf v}^{\prime}-{\bf v} 
_{w}\right))\left| \widehat{{\bf n}}\cdot \left( {\bf v}^{\prime}-{\bf v} 
_{w}\right) \right| f_{i}^{(1)}({\bf r},{\bf v}^{\prime};t)  \nonumber \\ 
& &
\left. -\Theta (-\widehat{{\bf n}}\cdot \left({\bf v}_{1}-{\bf v} 
_{w}\right) )\left| \widehat{{\bf n}}\cdot \left( {\bf v}_{1}-{\bf v} 
_{w}\right) \right| f_{i}^{(1)}({\bf r},{\bf v}_{1};t)\right\}  \label{a.7} 
\end{eqnarray} 

The change in the kinetic energy due to the boundary conditions is  
\begin{eqnarray}
\label{a.8} 
\int d{\bf r}\int d{\bf v}_{1}
\frac{1}{2}m_{i}v_{1}^{2}T_{Wi}f_{i}^{(1)}({\bf r},{\bf v} 
_{1};t)&=&\int_{\Xi }d{\bf s}
\int d{\bf v}_1\left[e_{i}^{\prime}\left( 
v_1\right)-e_{i}(v_1)\right] \left| \widehat{{\bf n}}\cdot \left( {\bf 
v}_1-{\bf v}_{w}\right)\right| \nonumber\\
& & \times
\Theta (-\widehat{{\bf n}}\cdot \left({\bf v}_1-{\bf v}_{w}\right))
f_{i}^{(1)}({\bf s},{\bf v}_1;t),
\end{eqnarray}
where $e_{i}(v_1)$ and $e_{i}^{\prime}(v_1)$ are the kinetic energies for 
particles coming into and leaving the surface. They are given by   
\begin{equation}
\label{a.8bis}
e_{i}(v_1)=\frac{1}{2}m_{i}v_1^{2},
\end{equation}
\begin{eqnarray}
\label{a8bbis} 
e_{i}^{\prime}(v_1) &=&\int d{\bf v}\Theta (\widehat{{\bf n}} 
\cdot \left( {\bf v}-{\bf v}_{w}\right) )\frac{1}{2}m_{i}v^{2}\delta 
\left({\bf v}-{\bf v}_1+2\widehat{{\bf n}}\cdot \left({\bf v}_1-{\bf v} 
_{w}\right) \widehat{{\bf n}}\right) \\ 
&=&\frac{1}{2}m_{i}\left({\bf v}_1-2\left(\widehat{{\bf n}}\cdot {\bf v}_1- 
v_{w}\right) \widehat{{\bf n}}\right) ^{2}\Theta (-\widehat{{\bf n}}\cdot 
\left({\bf v}_1-{\bf v}_{w}\right)). 
\end{eqnarray} 
Thus, one gets 
\begin{equation} 
e_{i}^{\prime}(v_1) -e_{i}(v_1)=2m_{i}v_{w}^{2}\left( 1-
\frac{\widehat{{\bf n}}\cdot {\bf v}_1}{v_{w}}\right)  
\label{a.9} 
\end{equation}
Here, $v_{w}$ is the speed of the wall, taken along the normal 
$\widehat{{\bf n}}$. The rate of energy change due to the wall becomes  
\begin{eqnarray} 
\int d{\bf r}\int d{\bf 
v}_{1}\frac{1}{2}m_{i}v_{1}^{2}T_{Wi}f_{i}^{(1)}({\bf r},{\bf 
v}_{1};t)&=&2m_{i}v_{w}^{3}\int d{\bf v}_1
\left(1-\frac{\widehat{{\bf n}}\cdot {\bf v}_1}{v_{w}}\right)^{2}
\nonumber\\
& & \times
\int_{\Xi}d{\bf s}
\Theta (-\widehat{{\bf n}}\cdot \left({\bf v}_1-{\bf v}_{w}\right) 
)f_{i}^{(1)}({\bf s},{\bf v}_1;t)  \label{a.10} 
\end{eqnarray} 
If the wall speed is much larger than other characteristic velocities,  
$1-\widehat{{\bf n}}\cdot {\bf v}/v_w\rightarrow 1$, and so Eq.\ (\ref{a.10})
becomes approximately  
\begin{equation} 
\int d{\bf r}\int d{\bf 
v}_{1}\frac{1}{2}m_{i}v_{1}^{2}T_{Wi}f_{i}^{(1)}({\bf r},{\bf v} 
_{1};t)\rightarrow m_{i}v_{w}^{3}n_{i}A,  \label{a.11} 
\end{equation} 
where $A$ is the area of the wall and $n_{i}$ is the density of incident 
particles of species $i$. The rate of change of the temperature is then  
\begin{equation} 
\frac{\partial T_{i}}{\partial t}=\frac{2A}{3V}m_{i}v_{w}^{3}-\zeta _{i}T_{i} 
\label{a.12} 
\end{equation} 
which gives the steady state condition (\ref{5.3}). 
 
Another estimate is given by representing the incident distribution as a 
Gaussian 
\begin{eqnarray}
\label{a.13}
2n_{i}Am_{i}v_{w}^{3}\frac{1}{\sqrt{\pi}}& & \int_{-\infty 
}^{v_{w}/v_{0i}}dv_{x}^{\ast }\left( 1-\frac{v_{1i}}{v_{w}}v_{x}^{\ast 
}\right) ^{2}e^{-\left( v^{\ast }+v_{2}^{\ast }\right) ^{2}}  
=2n_{i}Am_{i}v_{w}v_{1i}^{2}\nonumber\\
& & \times
\left( \frac{1}{2\sqrt{\pi }}\left( 
v_{w}^{\ast }+v_{2}^{\ast }\right) e^{-\left( v_{w}^{\ast }+v_{2}^{\ast 
}\right) ^{2}}+\frac{1}{2}\left[ \left( v_{w}^{\ast }+v_{2}^{\ast }\right) 
^{2}+2\right] \left( \mathop{\rm erf} \left( v_{w}^{\ast }-v_{2}^{\ast 
}\right) +1\right) \right)  
\end{eqnarray} 
Here $v_{w}^{\ast }=v_{w}/v_{1i},$ $v_{1i}^{2}\allowbreak =2T_{1i}/m_{i}$ 
characterizes the temperature of the incident particles and $%
v_{2}=v_{2}^{\ast }v_{1i}$ characterizes their mean speed toward the wall. 
For large $v_{w}^{\ast }+v_{2}^{\ast }$ the condition (\ref{a.12}) is 
recovered, while for small $v_{w}^{\ast }+v_{2}^{\ast }$ $\ $it becomes%
\begin{equation} 
\frac{\partial T_{i}}{\partial t}=\frac{4A}{3V}v_{w}T_{1i}-\zeta _{i}T_{i} 
\label{a.14} 
\end{equation} 
If $T_{1i}\approx T_{i}$ then the steady state condition for the HCS, 
constant cooling rate, is recovered.

\begin{figure}[tbp] 
\caption{Time evolution of $\protect\gamma(t)=T_1(t)/T_2(t)$ for $\protect%
\phi=0.1$, $\protect\sigma_1/\protect\sigma_2=\protect\phi_1/\protect\phi%
_2=1 $, $m_1/m_2=8$ and two values of $\protect\alpha$: $\protect\alpha=0.8$ 
and $\protect\alpha=1$. } 
\label{fig1} 
\end{figure} 
 
\begin{figure}[tbp] 
\caption{Plot of the temperature ratio $T_1/T_2$ as a function of the mass 
ratio $m_1/m_2$ for $\protect\sigma_1/\protect\sigma_2=\protect\phi_1/%
\protect\phi_2=1$, and two different values of $\protect\alpha$: $\protect%
\alpha=0.95$ (solid line and circles) and $\protect\alpha=0.8$ (solid line 
and triangles). The lines are the Enskog predictions and the symbols refer 
to the MD simulation results. The open (solid) symbols correspond to $%
\protect\phi=0.1$ ($\protect\phi=0.2$). } 
\label{fig2} 
\end{figure} 
 
\begin{figure}[tbp] 
\caption{Plot of the temperature ratio $T_1/T_2$ as a function of the size 
ratio $\protect\sigma_1/\protect\sigma_2$ for $m_1/m_2=\protect\phi_1/%
\protect\phi_2=1$, and two different values of $\protect\alpha$: $\protect%
\alpha=0.95$ (lines and circles) and $\protect\alpha=0.8$ (lines and 
triangles). The lines are the Enskog predictions and the symbols refer to 
the MD simulation results. The solid (dashed) lines correspond to $\protect%
\phi=0.1$ ($\protect\phi=0.2$) while the open (solid) symbols correspond to $%
\protect\phi=0.1$ ($\protect\phi=0.2$). } 
\label{fig3} 
\end{figure} 
 
\begin{figure}[tbp] 
\caption{Plot of the temperature ratio $T_1/T_2$ as a function of 
composition $\protect\phi_1/\protect\phi_2$ for $m_1/m_2=8$, $\protect\sigma%
_1/\protect\sigma_2=2$, and two different values of $\protect\alpha$: $%
\protect\alpha=0.95$ (lines and circles) and $\protect\alpha=0.8$ (lines and 
triangles). The lines are the Enskog predictions and the symbols refer to 
the MD simulation results. The solid (dashed) lines correspond to $\protect%
\phi=0.1$ ($\protect\phi=0.2$) while the open (solid) symbols correspond to $%
\protect\phi=0.1$ ($\protect\phi=0.2$). } 
\label{fig4} 
\end{figure} 
 
\begin{figure}[tbp] 
\caption{Plot of the temperature ratio $T_1/T_2$ as a function of the mass 
ratio $m_1/m_2$ for $\protect\phi=0$, $\protect\sigma_1/\protect\sigma_2=%
\protect\phi_1/\protect\phi_2=1$, and two different values of $\protect%
\alpha $: $\protect\alpha=0.8$ and $\protect\alpha=0.6$. The solid lines are 
the results for the HCS while the dashed lines are the results for the 
driven steady state achieved from the stochastic thermostat. } 
\label{fig5} 
\end{figure} 
 
\begin{figure}[tbp] 
\caption{Plot of the temperature ratio $T_1/T_2$ as a function of 
composition $\protect\phi_1/\protect\phi_2$ for $\protect\phi=0$, $m_1/m_2=2$%
, $\protect\sigma_1/\protect\sigma_2=1$, and two different values of $%
\protect\alpha$: $\protect\alpha=0.8$ and $\protect\alpha=0.6$. The solid 
lines are the results for the HCS while the dashed lines are the results for 
the driven steady state achieved from the stochastic thermostat. } 
\label{fig6} 
\end{figure} 
 
\begin{figure}[tbp] 
\caption{Plot of the relative temperature ratio $\protect\gamma(\protect%
\alpha,\protect\phi)/\protect\gamma(\protect\alpha,0)$ as a function of the 
total solid volume fraction $\protect\phi$ for $m_1/m_2=\protect\sigma_1/%
\protect\sigma_2=2$, $\protect\phi_1/\protect\phi_2=1$, and two different 
values of $\protect\alpha$: $\protect\alpha=0.8$ and $\protect\alpha=0.6$. 
The solid lines are the results for the HCS while the dashed lines are the 
results for the driven steady state achieved from the stochastic thermostat.  
} 
\label{fig7} 
\end{figure} 
 

\begin{references} 

\bibitem{GS95} A. Goldshtein and M. Shapiro, J. Fluid Mech. {\bf 282}, 75 
(1995); T. P. C. van Noije and M. H. Ernst, Gran. Matt. {\bf 1}, 57 (1998). 
 
\bibitem{BRC96} J. J. Brey, M. J. Ruiz-Montero, and D. Cubero, Phys. Rev. E  
{\bf 54}, 3664 (1997). 
 
\bibitem{HOB00} M. Huthman, J. Orza, and R. Brito, Gran. Matt. {\bf 2}, 189 
(2000). 
 
\bibitem{GD99} V. Garz\'{o} and J. W. Dufty, Phys. Rev. E {\bf 60}, 5706 
(1999). 
 
\bibitem{MG02} J. M. Montanero and V. Garz\'{o}, Gran. Matt. {\bf 4}, 17 
(2002). 
 

\bibitem{SD01} A. Santos and J. Dufty, Phys. Rev. Lett. {\bf 86}, 4823 
(2001); Phys. Rev. E {\bf 64}, 51305 (2001). 
 
\bibitem{L01} J. F. Lutsko, Phys. Rev. E {\bf 63}, 061211 (2001); 
 
\bibitem{D02} J. W. Dufty, J. J. Brey, and J. Lutsko, Phys. 
Rev. E {\bf 65}, 051303 (2002); J. Lutsko, J. J. Brey, and J. W. Dufty, 
Phys. Rev. E {\bf 65}, 051304 (2002). 
 
\bibitem{BT02} A. Barrat and E. Trizac, Gran. Matt. {\bf 4}, 57 (2002).
 
\bibitem{FM01} K. Feitosa and N. Menon, Phys. Rev. Lett. {\bf 88}, 198301 
(2002).
 
\bibitem{WP02} R. Wildman and D. Parker, Phys. Rev. Lett. {\bf 88}, 064301 
(2002). 
 
\bibitem{KH00}A. Karion and M. L. Hunt, Powder Tech. {\bf 109}, 145 (2000). 

\bibitem{CH02}R. Clelland and C. Hrenya, Phys. Rev. E {\bf 65}, 031301 
(2002).

\bibitem{BDS97} J. J. Brey, J. W. Dufty, and A. Santos, J. Stat. Phys. {\bf %
87}, 1051 (1997). 
 
\bibitem{DG01} J. W. Dufty and V. Garz\'o, J. Stat. Phys. {\bf 105}, 
723 (2001). 

\bibitem{GH72} E. W. Grundke and D. Henderson, Mol. Phys. {\bf 24}, 269 
(1972); L. L. Lee and D. Levesque, Mol. Phys. {\bf 26}, 1351 (1973). 
 
\bibitem{CS}See for instance, D. H. L. Yan, K. Y. Chan, and D. 
Henderson, Mol. Phys. {\bf 88}, 1237 (1996); D. V. Matyushov and B. M. 
Landanyi, J. Chem. Phys. {\bf 107}, 5815 (1997); D. Cao, K. Y. Chan, D. 
Henderson, and W. Wang, Mol. Phys. {\bf 98}, 619 (2000); A. Santos, S. B. 
Yuste, and M. L\'opez de Haro, J. Chem. Phys. (to be 
published) and cond-mat/0203182. 
 
\bibitem{MP02} U. Marini Bettolo Marconi and A. Puglisi, Phys. 
Rev. E {\bf 65}, 051305 (2002).
 

\bibitem{footnote} Haff's law is a necessary, but not sufficient, 
condition for the scaling of the HCS. In principle there could be a locally 
inhomogeneous state such that the total energy is the sum of the thermal 
energy plus the kinetic energy of convective flows. It is possible that both 
contributions could decay with the same power law as in Haff's law, but 
with a different cooling rate $\zeta$. In fact, there is some indication 
that this may occur in the late stages of instability for the HCS (J. Wakou, 
R. Brito, and M. H. Ernst, cond-mat/0103086). We assume that in our case the 
initial state is always HCS and that a transition interval would be seen in 
any possible evolution to a different Haff's law. The onset of this 
transition interval is what has been identified here as the violation of 
Haff's law.


\bibitem{thermostat} T. P. C. van Noije and M. H. Ernst, Gran. Matt. {\bf 
1}, 57 (1998). 

\bibitem{EM90} D. J. Evans and G.P. Morriss, {\em Statistical Mechanics of 
Nonequilibrium Liquids} (Academic Press, London, 1990). 
 
\bibitem{DM96} D. Williams and F. MacKintosh, Phys. Rev. E {\bf 54}, R9 
(1996). 
 
\bibitem{HBB00} C. Henrique, G. Batrouni, and D. Bideau, Phys. Rev. E {\bf 63%
}, 011304 (2000). 
 

\bibitem{BRM00} J. J. Brey, M. Ruiz-Montero, and F. Moreno, Phys. Rev. E  
{\bf 62}, 5339 (2000). 
 
\bibitem{GD02} V. Garz\'{o} and J. W. Dufty, Phys. Fluids {\bf 14}, 1476 
(2002). 


\bibitem{DvB77} J. R. Dorfman and H. van Beijeren in {\em Statistical 
Mechanics, \ Part B}, edited by B. Berne, (Plenum, NY, 1977). 

\end{references}
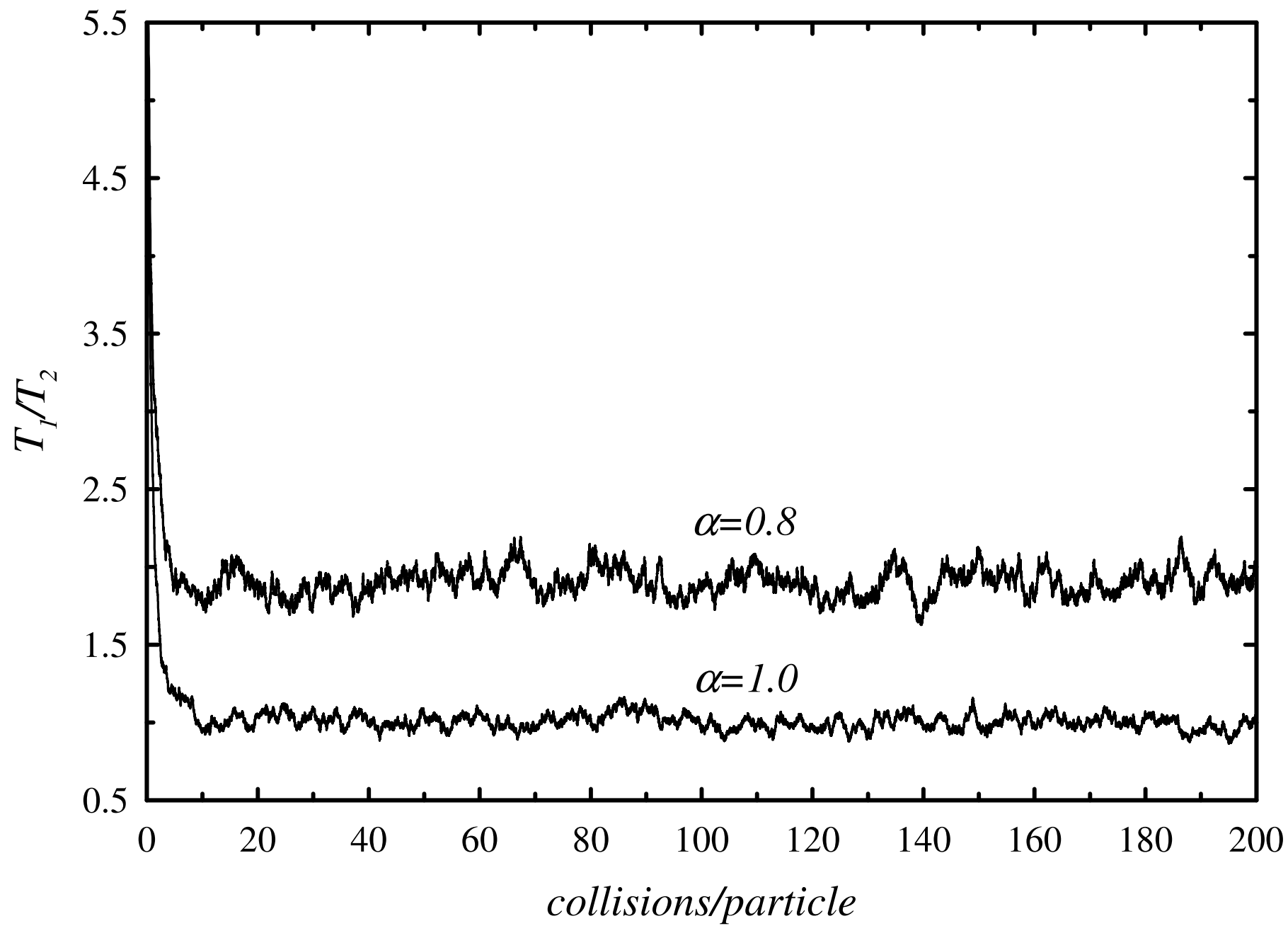
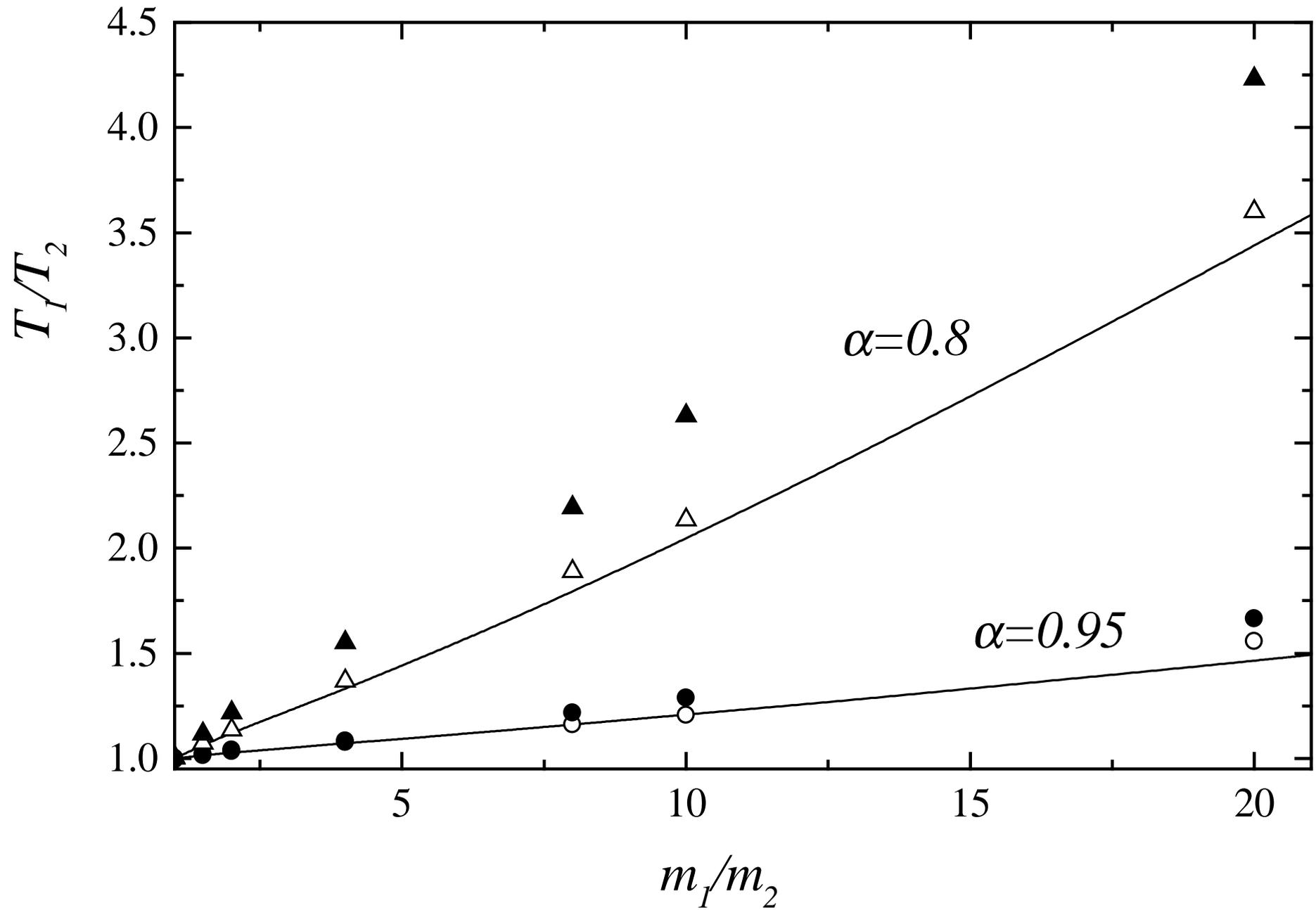
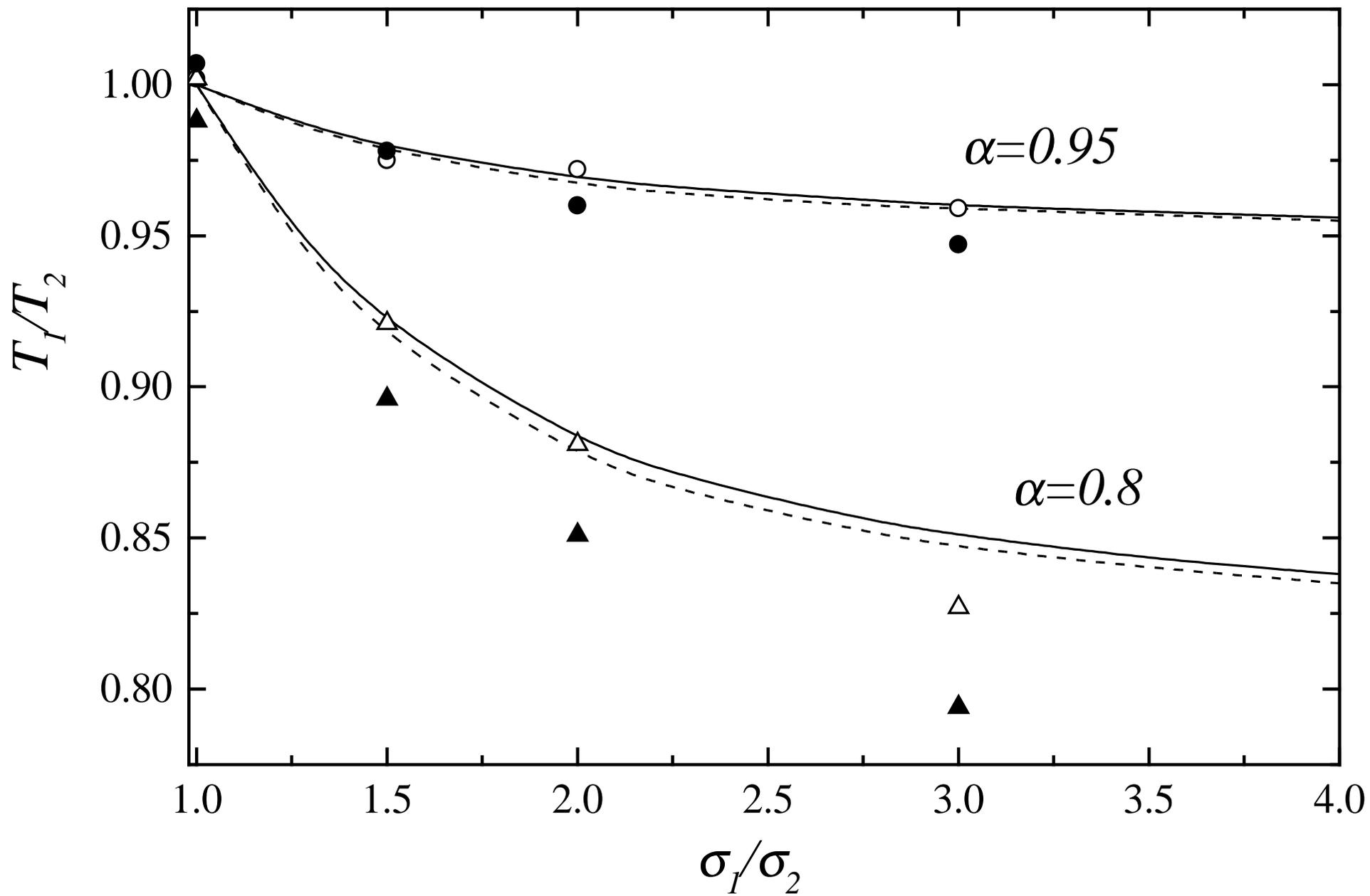
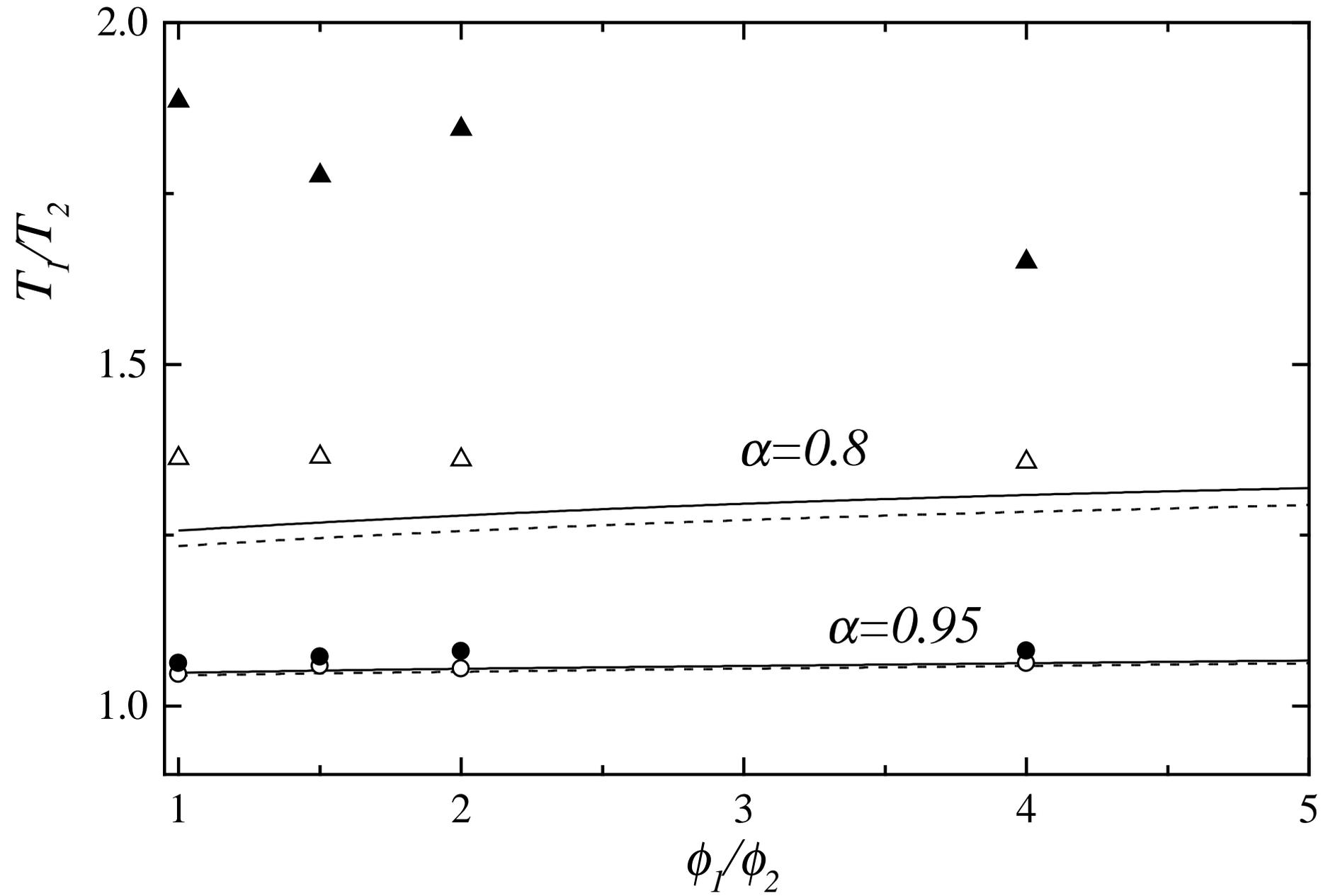
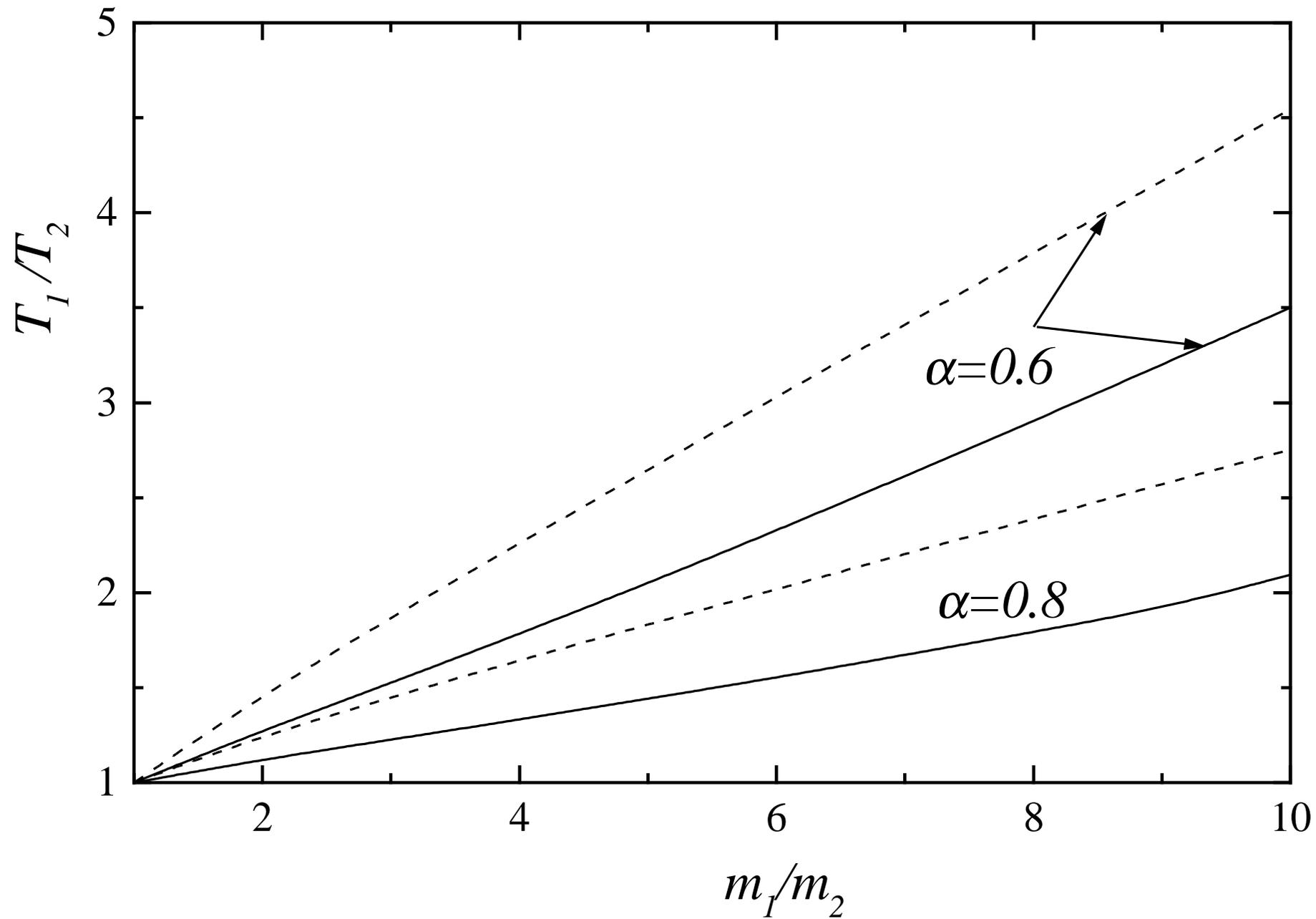
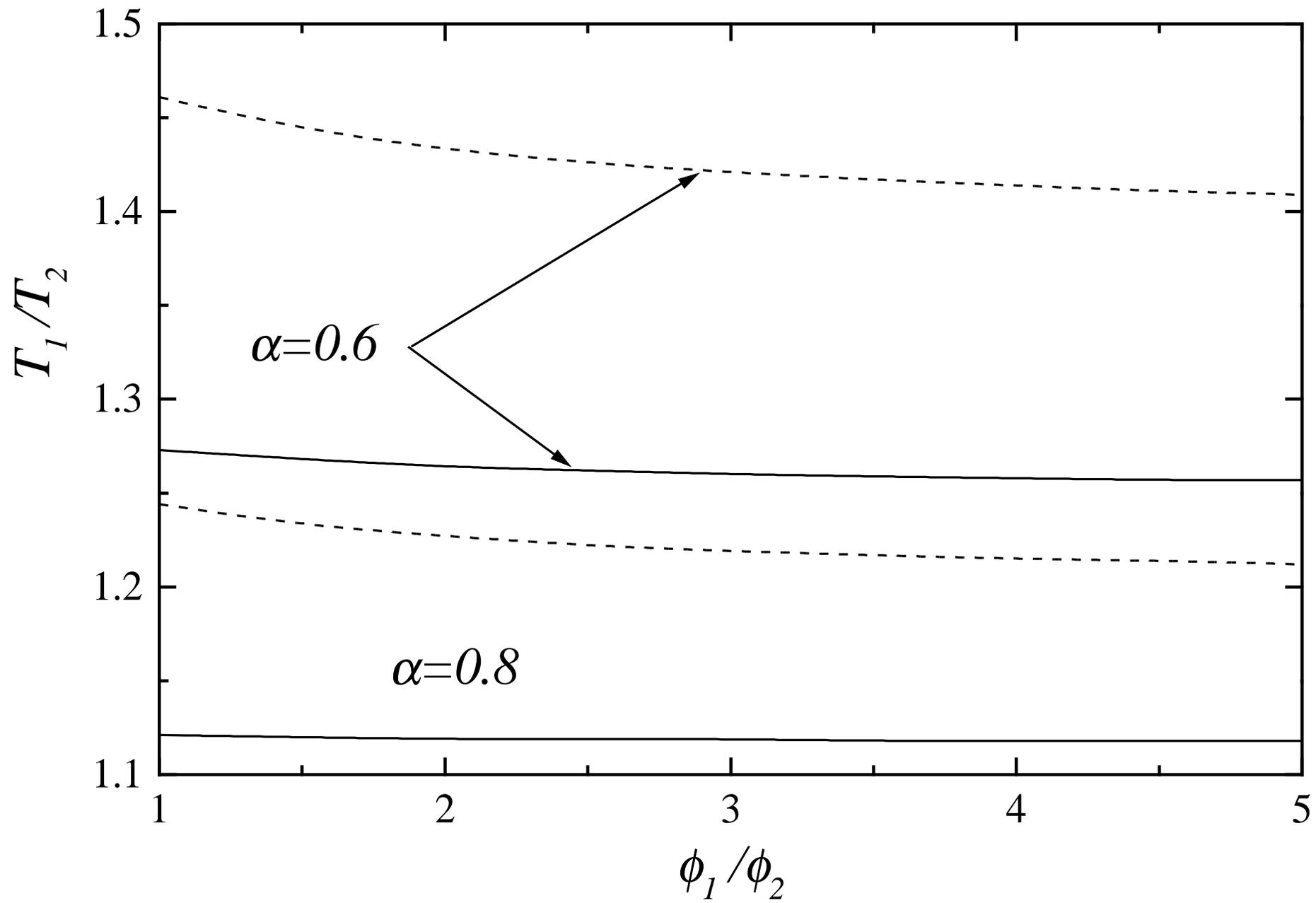
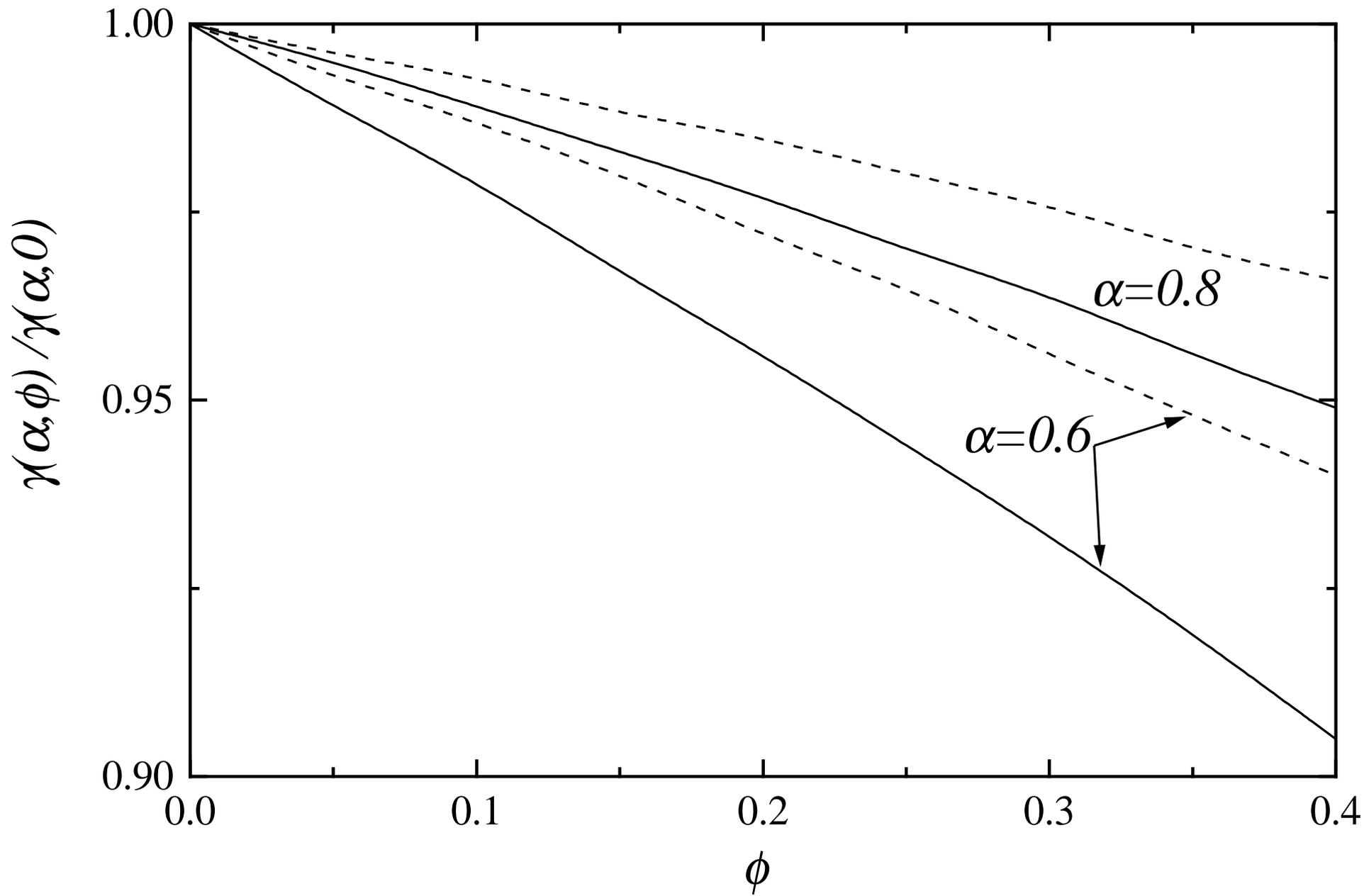
\end{document}